\documentclass[fleqn,usenatbib]{mnras}

\usepackage{newtxtext,newtxmath}
\usepackage{subcaption}
\usepackage[T1]{fontenc}

\usepackage{color}
\usepackage[dvipsnames]{xcolor}
\usepackage[normalem]{ulem}

\usepackage{graphicx}	
\usepackage{amsmath}	
\usepackage{float}
\usepackage{tabularx}
\usepackage{calc}

\newcommand{\bcdot}{\boldsymbol{\cdot}}
\newcommand{\bnabla}{\boldsymbol{\nabla}}

\newcolumntype{P}[1]{>{\centering\arraybackslash}p{#1}}

\title[Energy \& entropy schemes for CR hydrodynamics]{Comparing energy and entropy formulations for cosmic ray hydrodynamics}

\author[M. Weber et al.]{
Matthias Weber,$^{1, 2}$\thanks{E-mail: maweber@aip.de}
Timon Thomas,$^{1, 2}$
Christoph Pfrommer$^{1}$
\\
$^{1}$Leibniz-Institute for Astrophysics Potsdam (AIP), An der Sternwarte 16, 14482 Potsdam, Germany\\
$^{2}$Institut für Physik und Astronomie, Universität Potsdam, Karl-Liebknecht-Str. 24/25, 14476 Golm, Germany\\
}

\date{Accepted XXX. Received YYY; in original form ZZZ}

\pubyear{2022}

\begin{document}
\label{firstpage}
\pagerange{\pageref{firstpage}--\pageref{lastpage}}
\maketitle

\begin{abstract}
Cosmic rays (CRs) play an important role in many astrophysical systems. Acting on plasma scales to galactic environments, CRs are usually modeled as a fluid, using the CR energy density as the evolving quantity. This method comes with the flaw that the corresponding CR evolution equation is not in conservative form as it contains an adiabatic source term that couples CRs to the thermal gas. In the absence of non-adiabatic changes, instead evolving the CR entropy density is a physically equivalent option that avoids this potential numerical inconsistency. In this work, we study both approaches for evolving CRs in the context of magneto-hydrodynamic (MHD) simulations using the massively parallel moving-mesh code \textsc{Arepo}. We investigate the performance of both methods in a sequence of shock-tube tests with various resolutions and shock Mach numbers. We find that the entropy-conserving scheme performs best for the idealized case of purely adiabatic CRs across the shock while both approaches yield similar results at lower resolution. In this setup, both schemes operate well and almost independently of the shock Mach number. Taking active CR acceleration at the shock into account, the energy-based method proves to be numerically much more stable and significantly more accurate in determining the shock velocity, in particular at low resolution, which is more typical for astrophysical large-scale simulations. For a more realistic application, we simulate the formation of several isolated galaxies at different halo masses and find that both numerical methods yield almost identical results with differences far below common astrophysical uncertainties.
\end{abstract}





\section{Introduction}
\label{sec:introduction}
CRs represent the non-thermal particle population of an astrophysical plasma and arguably play a crucial role in understanding the self-regulated feedback mechanisms that are at work in galaxies and galaxy clusters \citep{Zweibel2017}. They acquire their high energies by diffusive acceleration processes at shocks  \citep{Marcowith2016} driven by supernovae (SNe), or by the relativistic energy feedback from active galactic nuclei \citep{Guo2008,Jacob2017a,Jacob2017b}. Concurrently, CRs suffer non-adiabatic cooling due to radiative and Coulomb losses, scattering off of self-excited magnetic fluctuations \citep{Kulsrud1969,Shalaby2021} and hadronic collisions. While energetic CR electrons thereby cool rapidly to negligible energies, rendering them dynamically insignificant in astrophysical systems, the momentum-carrying CR protons have much longer cooling times in comparison to their leptonic counterparts or thermal gas. This results in an approximate equipartition of the thermal, magnetic and CR pressure in the mid-plane of the Milky Way \citep{Boulares1990}, thus making CRs a promising agent of galactic feedback processes.

In the past decades, various approaches have been employed to numerically model the impact of CRs in astrophysical simulations. CRs act on a large range of scales, from characteristic plasma scales to galaxies to galaxy clusters. To explore CR dynamics in these macroscopic systems, the only computationally tractable approach is to model CRs collectively as a fluid. Commonly, a one-moment formulation for the CR fluid is applied in hydrodynamic and MHD simulations \citep{Hanasz2003, Enslin2007, Jubelgas2008, Booth2013, Salem2014, Girichidis2014, Pakmor2016b, Pfrommer2017, Dubois2019}, meaning that only a single scalar quantity (CR energy density or number density) is evolved in time. This setup is well suited for modeling the CR transport mechanisms of advection and diffusion. However, when applying this method to CR streaming, numerical instabilities may occur due to unlimited flux values \citep{Sharma2009}. Hence, further improvements were made by developing a two-moment formulation, in which the energy and flux densities of CRs are computed separately \citep{Jiang2018, Thomas2019, Thomas2022, Chan2019, Thomas2021}. The above algorithms exclusively use a simple `grey' approach for CR spectra, neglecting the different effects of and on CRs at different energies. To address this shortcoming, some codes were elaborated to handle spectrally resolved simulations, either using additional tracer particles \citep{Vaidya2018, Winner2019} or by adding multiple momentum bins per hydro-cell covering a wide range of the CR spectrum \citep{Miniati2001, Yang2017, Girichidis2020, Ogrodnik2021, Hopkins2022}.

Each of the previous models uses a two-fluid approximation to describe the thermal gas and CRs individually. Usually, the time evolution of the CR energy density is added as an extra relation to the conventional set of hydrodynamic/MHD equations, which represent conservation laws for mass, momentum and energy. As a consequence, the CR energy density does not separately follow such a conservation law. Any formulation of the CR energy density equation includes an adiabatic source term (either $P_\rmn{cr} \bnabla \bcdot \mathbfit{u}\,$ or $\,\mathbfit{u}\bcdot\bnabla P_\rmn{cr}$) that couples the CRs to the thermal gas. This term needs to be calculated in an additional step, thus preventing the CR equation from adopting a conservative form. While this is not a problem for smooth flows, the presence of non-vanishing spatial derivatives could in principle be problematic for shocks because a sudden jump in density/velocity could give rise to a continuous accumulation of numerical errors. This problem of non-uniqueness of the CR energy density is extensively discussed in \citet{Gupta2021}.

To overcome this potential numerical flaw, alternative schemes have been developed to integrate CR physics into the simulations based on ideas by \citet{Ryu1993}. Here, rather than CR energy, a modified CR entropy density ($\rho K_\rmn{cr} = P_\rmn{cr} / \rho^{\gamma_\rmn{cr}-1}$) is used as the relevant quantity to describe the CR fluid \citep{Kudoh2016, Semenov2021}. This approach has the evident benefit that the CR equation is in conservative form so that Godunov-type solvers can be straightforwardly applied. However, this formulation is only valid in the absence of non-adiabatic changes, where entropy is conserved. This is neither the case for astrophysical shocks, in which CRs are accelerated, nor for radiative, hadronic, and Alfv\'en wave cooling. In such cases, one would have to switch back to the energy description. Furthermore, the unavoidable dependence on mass density can lead to an immediate impact of (numerical) density fluctuations on the entropy variable, particularly in regimes of low resolution, which is the default scenario in large-scale simulations. Moreover, CR energy is not explicitly conserved in such schemes.

In their detailed study, \citet{Gupta2021} state that solving the two-fluid equations across shocks generally -- regardless of the numerical method used -- yields unique results only when an additional CR sub-grid closure is assumed. Without using such an artificial closure, they recommend adopting the energy-based method, where the total energy and CR energy are evolved in an unsplit scheme and the source term is added as $P_\rmn{cr} \bnabla \bcdot \mathbfit{u}\,$, since this approach proves to be most stable in that case. Further they argue that the entropy-conserving scheme does not provide satisfactory results in simple stability tests. Additionally, they point out that assuming a constant CR entropy across shocks is not physically justified because CRs are accelerated at shocks. Another study on the differences of the energy-based method and the entropy-conserving scheme is provided by \citet{Semenov2021}. According to their results, the use of the energy-based method leads to spurious entropy generation at shocks due to the numerical inaccuracies described earlier. Moreover, they find that this error depends on the shock Mach number and the adiabatic indices of the two fluids, while the entropy-conserving scheme does not suffer from any of these inaccuracies. This led them to conclude that the entropy-conserving scheme is the preferred choice to model CR fluids.

In this work, we investigate the differences of the CR energy and CR entropy formulations for CR transport using simulations that are carried out with the moving-mesh code \textsc{Arepo}. This paper is organized as follows. In Section~\ref{sec:cosmicrayhydrodynamics} we introduce the basic equations of CR-MHD physics and present the different methods to integrate CRs, namely the energy-based method and the entropy-conserving scheme. In Section~\ref{sec:testproblems} we perform a sequence of idealized tests for both numerical methods and compare their performance, also in the context of moving and static grids. In Section~\ref{sec:isolatedmodelsofgalaxyformation} we apply both schemes to a more realistic astrophysical scenario and model the formation of isolated galaxies. In Section~\ref{sec:conclusion} we summarize our main findings and conclusions. We use Heaviside-Lorentz units throughout this work and write $\mathbfit{a}\mathbfit{b}$ for the dyadic product of vectors $\mathbfit{a}$ and $\mathbfit{b}$.

\section{Cosmic ray magneto-hydrodynamics}
\label{sec:cosmicrayhydrodynamics}

In this section, we discuss the competing energy and entropy formulations for CR transport and how they are coupled to the MHD equations. Additionally, we present the extension of the existing energy-conserving numerical schemes to the entropy-conserving formulation of the CR transport equations.

\subsection{CR energy and entropy schemes}

In general, various CR transport phenomena influence how CRs are distributed in space once they leave their sources. This includes (but is not limited to) CR streaming or diffusion along magnetic field lines \citep{Skilling1971,Zweibel2013}, transport induced by magnetic field line wandering \citep{Jokipii1966,Shalchi2007}, CR interactions with turbulence \citep{Shalchi2009, Yan2011}, and guiding center drifts \citep{Gombosi2004,Schlickeiser2010}. In one of the common approximations, CRs are assumed to be co-moving with the bulk flow of the thermal particles and all additional transport process along or across the magnetic field are neglected. In this case, the evolution equation for the CR energy density, $\varepsilon_\rmn{cr}$, reads as:
\begin{align}
   \label{eq:crenergy}
    \frac{\partial\varepsilon_\rmn{cr}}{\partial t} + \bnabla\bcdot(\varepsilon_\rmn{cr}\mathbfit{u}) &= -P_\rmn{cr}\bnabla\bcdot\mathbfit{u} + \Gamma_\rmn{cr} + \Lambda_\rmn{cr},
\end{align}
where $P_\rmn{cr} = (\gamma_\rmn{cr} - 1) \varepsilon_\rmn{cr}$ is the CR pressure, $\gamma_\rmn{cr} = 4/3$ is the adiabatic index of the CRs, $\mathbfit{u}$ is the mean velocity of the thermal gas, and non-adiabatic gain and loss processes of CR energy are represented by $\Gamma_\rmn{cr}$, $\Lambda_\rmn{cr}$. The term $\bnabla\bcdot(\varepsilon_\rmn{cr}\mathbfit{u})$ describes the advection of CR energy with the gas flow while the term $P_\rmn{cr}\bnabla\bcdot\mathbfit{u}$ on the right-hand side of this equation states that CR energy is subject to adiabatic changes. This adiabaticity of the CRs suggests the definition of an proxy for the CR entropy given by
\begin{equation}
    \label{eq:proxy}
    K_\rmn{cr} = P_\rmn{cr} / \rho^{\gamma_\rmn{cr}},
\end{equation}
where $\rho$ is the gas mass density. We call $K_\rmn{cr}$ the \textit{specific CR entropy} or \textit{CR entropy} for short. The evolution equation for $K_\rmn{cr}$ is:
\begin{equation}
   \label{eq:crentropy}
    \frac{\partial (\rho K_\rmn{cr})}{\partial t} + \bnabla\bcdot(\rho K_\rmn{cr}\mathbfit{u}) = \frac{\gamma_\rmn{cr}-1}{\rho^{\gamma_\rmn{cr}-1}}(\Gamma_\rmn{cr} + \Lambda_\rmn{cr}),
\end{equation}
and states that CR entropy is solely advected with the gas-flow and is a conserved quantity in the absence of any explicit gains or losses of CR energy. The CR energy density does not have this favorable property and is a non-conserved quantity because of the adiabatic term which cannot be cast into a total-flux divergence form. This difference between the energy and entropy formulation for CR transport also influences the design of numerical schemes that implement these equations. While standard finite-volume schemes can be readily applied to the entropy equation \eqref{eq:crentropy}, these schemes cannot be directly applied to the adiabatic term of the CR energy equation \eqref{eq:crenergy} and other discretizations need to be made \citep{Kudoh2016, Gupta2021}.

CR exert forces on the thermal particles and are thus represented  through their pressure in the momentum and energy equations in the MHD system of equations
\begin{align}
    \label{eq:mass}
    \frac{\partial\rho}{\partial t} + \bnabla\bcdot(\rho\mathbfit{u}) &= 0,\\
    \label{eq:momentum}
    \frac{\partial(\rho\mathbfit{u})}{\partial t} + \bnabla\bcdot(\rho\mathbfit{uu}+P_\rmn{tot}\mathbf{1}+\mathbfit{BB}) &= \mathbf{0},  \\
    \label{eq:thermalenergy}
    \frac{\partial\varepsilon}{\partial t} + \bnabla\bcdot[(\varepsilon+P_\rmn{tot})\mathbfit{u}+\mathbfit{B}(\mathbfit{u}\bcdot\mathbfit{B})] &= P_\rmn{cr}\bnabla\bcdot\mathbfit{u} + \Gamma_\rmn{th} + \Lambda_\rmn{th},\\
    \label{eq:bfield}
    \frac{\partial\mathbfit{B}}{\partial t} + \bnabla\bcdot(\mathbfit{B}\mathbfit{u} + \mathbfit{u}\mathbfit{B}) &= 0,
\end{align}
where $\mathbfit{B}$ is the magnetic field, $\Gamma_\rmn{th}$ and $\Lambda_\rmn{th}$ are heating and cooling terms affecting the thermal energy density $\varepsilon_\rmn{th}$, $\varepsilon$ is the total MHD energy density given by 
\begin{equation}
    \varepsilon = \frac{\rho}{2}\mathbfit{u}^2 + \varepsilon_\rmn{th} + \frac{\mathbfit{B}^2}{2},
\end{equation}
and $P_\rmn{tot}$ is the total pressure of the composite fluid of CRs, thermal gas, and magnetic field, and is given by
\begin{equation}
    P_\rmn{tot} = P_\rmn{th} + P_\rmn{cr} + \frac{\mathbfit{B}^2}{2}.
\end{equation}
Similar to the CRs, thermal energy density and thermal pressure are linked by an equation of state:
\begin{align}
    \label{eq:eosth}
    P_\rmn{th} = (\gamma_\rmn{th}-1)\varepsilon_\rmn{th}, \,\,\,\,\,\, \text{where} \,\,\,\,\,\,\gamma_\rmn{th}=5/3.
\end{align}
Note that the total energy is conserved in the combined set of MHD equations together with the CR energy equation \eqref{eq:crenergy}, in the absence of explicit sources or sinks of CR or thermal energy. This cannot be guaranteed if the CR entropy equation \eqref{eq:crentropy} is used and thus energy errors will inevitably build up in simulations that employ this formulation for CR transport. Hence, the decision between the CR energy and entropy formulation is also a decision which conservation property is regarded to be more valuable. A priori, neither of them is more favorable.

In \citet{Pfrommer2017}, we detail our numerical scheme for integrating the CR energy equation and the modifications of the MHD scheme of \citet{Pakmor2016} to account for the additional CR pressure. We also implemented the CR entropy equation into the \textsc{Arepo} code. Because the CR entropy equation resembles the evolution equation of an advected and conserved scalar quantity, we use the routines of the \textsc{Arepo} code that integrate such conserved scalars to evolve the CR entropy. We regularly transform from the CR entropy to CR pressure in the code to use existing code structures for both the CR energy and entropy formalism. We keep the modifications to the code minimal in order to ensure that the details of the numerical scheme do not influence the simulations more then the choice of the CR transport formalism itself. For example, no changes to the Riemann solver, to the time-extrapolations of the MHD momentum or the thermal pressure are made, and both schemes use the same code that is based on the CR pressure.

\subsection{Shock detection and CR acceleration}
\label{sub:shockfinder}
Kinetic gas energy is dissipated into thermal energy at a shock. Diffusive shock acceleration and other plasma-physical processes can convert a fraction of the dissipated energy into energy contained in CRs. We model this conversion by subtracting parts of the dissipated energy and adding the same amount to the CR energy for computational cells that form the immediate downstream region of a shock. Details on the numerical algorithm that implements this conversion can be found in \citet{Pfrommer2017}. This existing injection algorithm is extended to be applicable with the CR entropy formalism. We first calculate the preexisting CR energy from the current value of the CR entropy, add the injected CR energy, and then recalculate the CR entropy from the updated value of the CR energy. This ensures energy conservation of dissipated energy during the injection procedure.

To model the injection of freshly accelerated CRs at the shock front, we employ the shock finding method developed by \citet{Schaal2015} and extended by \citet{Pfrommer2017}. We summarize the main points of this algorithm for completeness here. The shock finding algorithm identifies a shock zone by applying the following local cell-based criteria:
\begin{enumerate}
    \item $\nabla\bcdot\mathbfit{u} < 0$,
    \item $\nabla \Tilde{T} \bcdot \nabla\rho > 0$,
    \item $\Tilde{\mathcal{M}} > \Tilde{\mathcal{M}}_\rmn{min}$,
\end{enumerate}
where $\Tilde{\mathcal{M}}$ is the (numerically stabilized) shock Mach number and $\Tilde{T}$ is the pseudo temperature of the composite gas, defined via
\begin{equation}
    k_\rmn{B}\Tilde{T} = \frac{P}{n} = \frac{\mu m_\rmn{p}(P_\rmn{th}+P_\rmn{cr})}{\rho},
\end{equation}
where $n$ is the gas number density, $m_\rmn{p}$ is the proton rest mass, $\mu$ is the mean molecular weight, and $k_\rmn{B}$ denotes the Boltzmann constant. Criterion (i) detects converging flows, which is the essential condition for the presence of a shock. To filter spurious shocks such as tangential or contact discontinuities, criterion (ii) is applied. These discontinuities are characterized by constant pressure across their surfaces which implies that the temperature and density change in opposite directions and therefore the corresponding gradients have different signs. Criterion (iii) gives a minimum threshold for the Mach number to distinguish numerical noise from physical shocks, which we chose to be $\Tilde{\mathcal{M}}_\rmn{min}=1.3$ in this work.

\section{Test problems}
\label{sec:testproblems}
In this section, we perform a suite of test problems to compare the performance of both methods described in Section~\ref{sec:cosmicrayhydrodynamics}. By default, all simulations in this section are performed with the moving-mesh setup of \textsc{Arepo} using standard parameters for mesh regularisation \citep{Vogelsberger2012, Pakmor2016, Weinberger2020} and a grid that is initially equally spaced.

\subsection{Pressure balance}
\label{subsec:PB}

In this first test, we set up a contact discontinuity characterized by a uniform density, a uniform total pressure but jumping CR and thermal pressures over the discontinuity. The gas is initially moving with a constant velocity inside a periodic simulation domain. Because the total pressure is constant, these initial conditions are dynamically stable in the sense that the CR and thermal pressure jumps at the contact discontinuity should not seed any additional motions. The resulting profiles for the gas density, velocity, thermal, and CR pressure should coincide with their respective initial values after each periodic crossing of the contact discontinuity through the simulation domain. This pressure balance test offers a simple way to test the basic stability of a numerical method in hydrodynamics. If a method fails this test, it is likely to fail even in more complex simulations. We use the same setup of \citet{Gupta2021}. The contact discontinuity is set up at $x=0.5$ inside a periodic simulation domain of length $L=1$. The initial conditions for the left and the right state are defined as $\{\rho, u, P_\rmn{th}, P_\rmn{cr}\}_L = \{1, 1, 0.1, 0.9 \}$ and $\{\rho, u, P_\rmn{th}, P_\rmn{cr}\}_R = \{1, 1, 0.9, 0.1 \}$. We use a resolution of $N=1000$ mesh cells and the moving-mesh setup of \textsc{Arepo}.

\begin{figure}
    \centering
    \includegraphics[width=\linewidth]{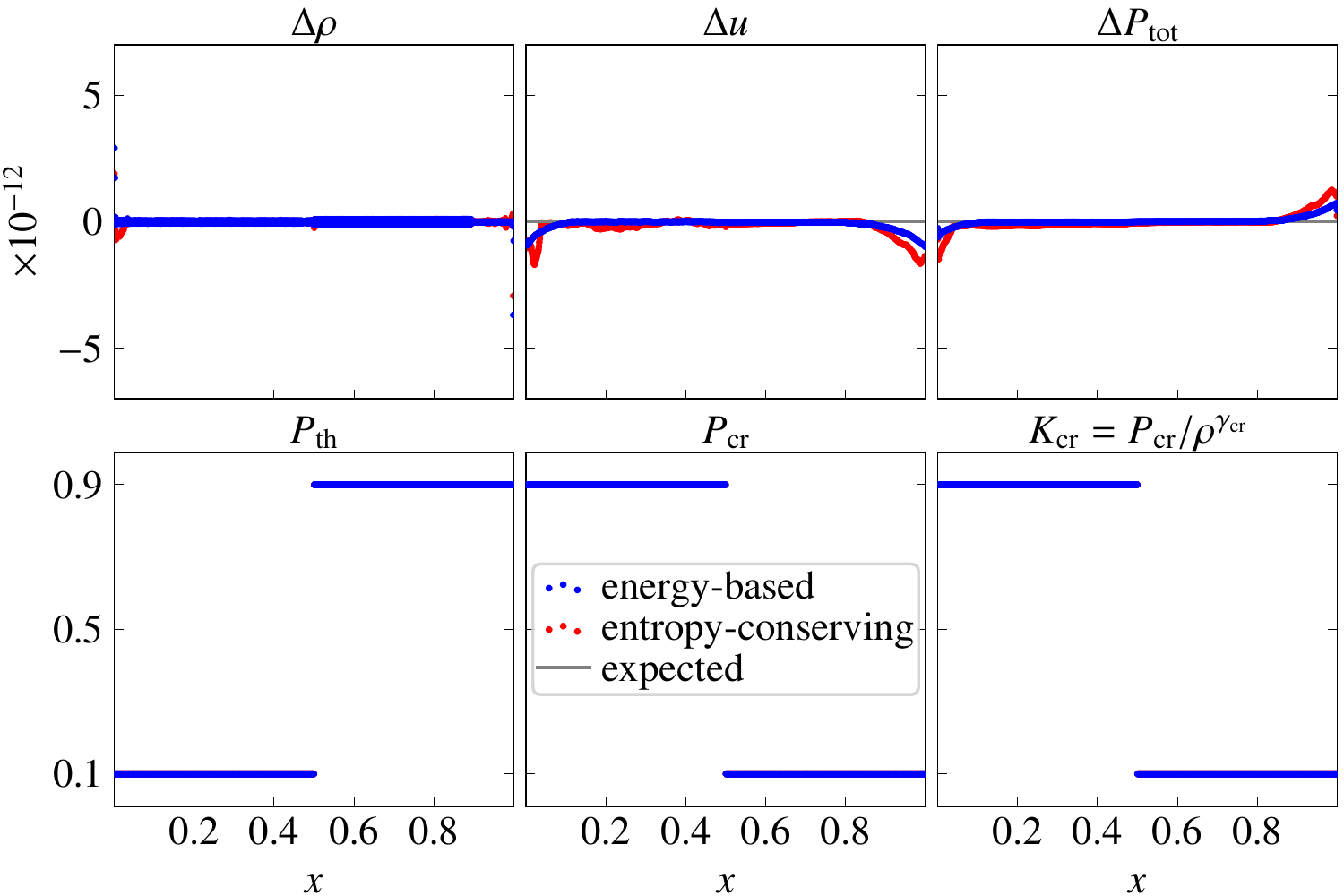}
    \caption{Results of the pressure balance test with periodic boundary conditions at $t=1.0$, i.e. after one box crossing time. In the top row we plot the deviations of the simulation results from the expected values, i.e. $\Delta \rho$, $\Delta u_x$ and $\Delta P_\rmn{tot}$. Note that the limits of the respective $y$-axis are set to $\Delta y\lesssim 5\times10^{-12}$ (where $y\in\{\rho,u,P_\rmn{tot}\}$). The bottom row shows quantities that are initially discontinuous across $x=0.5$, i.e. $P_\rmn{th}$, $P_\rmn{cr}$ and $K_\rmn{cr}$}.
    \label{fig:PB}
\end{figure}

Figure~\ref{fig:PB} shows the simulation results of the pressure balance test at $t=1$, i.e. after one box-crossing time. Note that the limits of the respective $y$-axis in the top row are set to $\Delta y\lesssim 5\times10^{-12}$. Minor blips form in the density, velocity and pressure profiles using either CR formulation. Because the blips have a low amplitude, they do not influence the overall dynamics. \citet{Gupta2021} performed the same test employing both the energy and entropy formalism for CR transport in the \textsc{PLUTO} code. They found that their numerical scheme produces deviations in the percentage regime for the simulation with the entropy formalism and that truncation errors in simulations with the energy formalism depended on details of the numerical algorithm they chose.

\subsection{Shock tubes}
\label{sub:shocktubes}
For our next test, we perform a sequence of one-dimensional (1D) shock-tube simulations with various Mach numbers $\mathcal{M} = u_\rmn{sh} / c_\rmn{s,\,pre}$, where $u_\rmn{sh}$ is the shock velocity in the lab frame and $c_\rmn{s,\,pre}$ is the pre-shock sound speed. We vary $\mathcal{M}$ from $1.5$ to $100$ and use several resolutions ranging from $N=30$ to $10^{4}$ mesh cells. The general setup to this problem is identical to the one presented by \citet{Pfrommer2017}. We set up a box of length $L=10$ containing a discontinuity at $x=5$. Gas in the left half-space ($x<5$) has a density of $\rho=1$ and a relative CR pressure of $X_\rmn{cr}=P_\rmn{cr}/P_\rmn{th} = 2$. We vary the thermal pressure and the CR pressure between different simulations in this region to achieve the desired Mach numbers of the shock while keeping the pressure ratio $X_\rmn{cr}$ constant. A shock tube forms because this half-space is initially over-pressurised with respect to the right half-space ($x>5$) that contains gas at a low density of $\rho = 0.125$. The thermal and CR pressures in this region are the same for all simulations and are set to $P_\rmn{th}=P_\rmn{cr}=0.05$. The fluid is initially at rest, $u_x=0$, and we use reflective boundary conditions. For the exact initial values we refer to Table~\ref{table:inits}. We perform two sets of simulations: one that only considers adiabatic changes of CRs (discussed in Section \ref{subsub:NoCRAcc}) and one that additionally accounts for non-adiabatic changes in the form of CR acceleration at the shock (discussed in Section \ref{subsub:CRAcc}).

\begin{table}
\caption{Initial conditions for the shock-tube tests with various Mach numbers $\mathcal{M}$. The indices L and R denote values of the left and right half-space, respectively.}
\centering
	\begin{tabular}{cccccccc}
	\hline\hline
	 $\mathcal{M}$ & $u_x$ & $\rho_\rmn{L}$ & $P_\rmn{th,\,L}$ & $X_\rmn{cr,\, L}$ & $\rho_\rmn{R}$ & $P_\rmn{th,\,R}$ & $X_\rmn{cr,\, R}$\\
	 \hline
    \multicolumn{8}{c}{\phantom{\Big|}Without CR shock acceleration:}\\
    \hline
	 1.5 & 0 & 1 & 0.24263 &2  &0.125 &0.05   & 1\\ 
	 2   & 0 & 1 & 0.54795 &2  &0.125 &0.05   & 1\\
	 3   & 0 & 1 & 1.4182  &2  &0.125 &0.05   & 1\\
	 5   & 0 & 1 & 4.1911  &2  &0.125 &0.05   & 1\\
	 10  & 0 & 1 & 17.172  &2  &0.125 &0.05   & 1\\
	 15  & 0 & 1 & 38.804  &2  &0.125 &0.05   & 1\\
	 30  & 0 & 1 & 155.61  &2  &0.125 &0.05   & 1\\
	 60  & 0 & 1 & 622.84  &2  &0.125 &0.05   & 1\\
	 100 & 0 & 1 & 1730.4  &2  &0.125 &0.05   & 1\\
	 \hline
    \multicolumn{8}{c}{\phantom{\Big|}With CR shock acceleration:}\\
	 \hline
	 9.56  & 0 & 1 & 17.172  &2  &0.125 &0.05   & 1\\
	 \hline
	\end{tabular}
\label{table:inits}	
\end{table}

\subsubsection{Adiabatic CRs}
\label{subsub:NoCRAcc}

\begin{figure}
\centering
    \includegraphics[width=\linewidth]{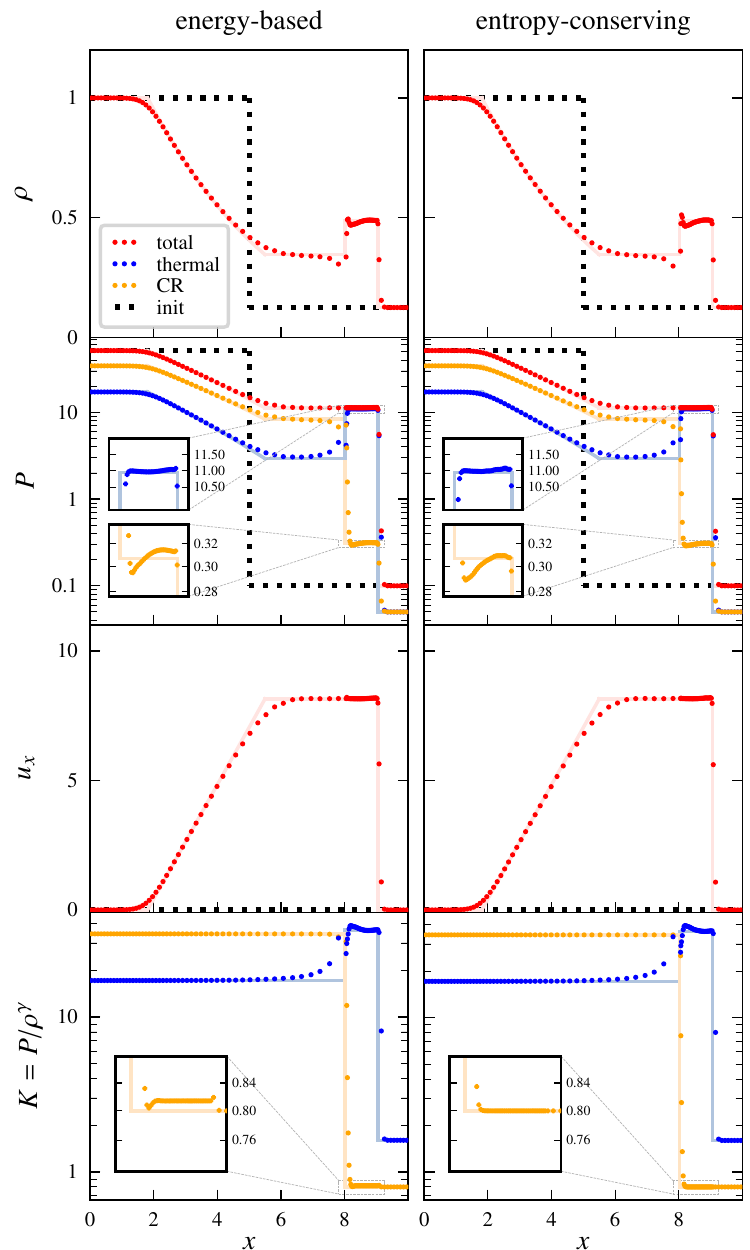}
    \caption{Shock-tube test for a composite of CRs and thermal gas while omitting CR acceleration at the shock. The left column displays the results using the energy-based method and the right column of the entropy-conserving scheme. Shown are 1D simulations with a resolution of $N = 100$ mesh cells and $\mathcal{M}=10$ at $t = 0.37$. We plot from top to bottom: mass density $\rho$, pressure $P$, velocity $u_x$ and entropy $P_i/\rho^{\gamma_i}$, where $i\in\{\rmn{cr},\rmn{th}\}$. Analytic solutions are shown as solid lines in semi-transparent colour, and simulation results as dots. The inset panels in the second and bottom row show magnifications of the corresponding post-shock regime, indicated by the dashed rectangles. }
    \label{fig:ShockTubeM10p100}
\end{figure}

Figure~\ref{fig:ShockTubeM10p100} shows the results of the 1D shock-tube test with $\mathcal{M}=10$ and only accounting for adiabatic changes of the CRs. The left-hand panel shows the results using the energy-based method, the right-hand panel shows the outcome using the entropy-conserving scheme. We perform both runs with identical initial conditions (see Table~\ref{table:inits}) and a spatial resolution of $N=100$ mesh cells. The simulation results resemble the well-known Sod-shock tube: a rarefaction develops to the left while a contact discontinuity and a shock form to the right of the initial discontinuity. Because the CRs evolve only adiabatically, the CR entropy is expected to be almost featureless. The only discontinuity in this profile should coincide with the contact discontinuity and separate the high CR-entropy gas from the low CR-entropy gas.  

As shown in Fig.~\ref{fig:ShockTubeM10p100}, both methods show nearly identical results and are in very good agreement with the analytic solutions (solid lines in semi-transparent colour; values adopted from \citealt{Pfrommer2006}) for density $\rho$, thermal pressure $P_{\rmn{th}}$, CR pressure $P_{\rmn{cr}}$, velocity $u_x$ and CR and thermal entropy, $K_{\rmn{cr}}$ and $K_\rmn{th}=P_\rmn{th}/\rho^{\rmn{\gamma_\rmn{th}}}$, where $\gamma_\rmn{th}=5/3$. To give a more detailed view, we zoom into the post-shock regime of the pressure and entropy plots as indicated by the inset panels in the second and bottom row. The magnified boxes show the post-shock region around at the analytical solution. We note that even at this magnification $P_{\rmn{th}}$ and $P_{\rmn{cr}}$ are still in good agreement with the analytical solution and deviate only about 1 per cent for both numerical schemes. A similar result is obtained for the entropy $K$. The entropy-conserving scheme does an excellent job of adiabatically compressing the CRs at the shock while keeping the CR entropy density constant across the shock. The energy-based method generates an artificial amount of CR entropy at the shock with a deviation from the analytic solution in the 2 per cent regime using our moving-mesh setup.

\begin{figure*}
    \includegraphics[width=0.49\linewidth]{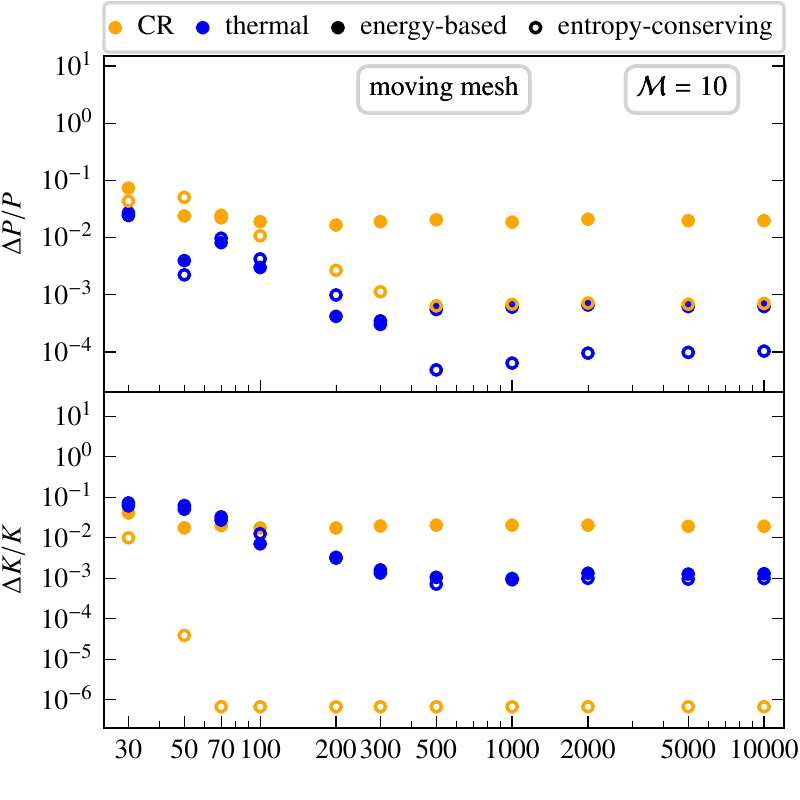}
    \includegraphics[width=0.49\linewidth]{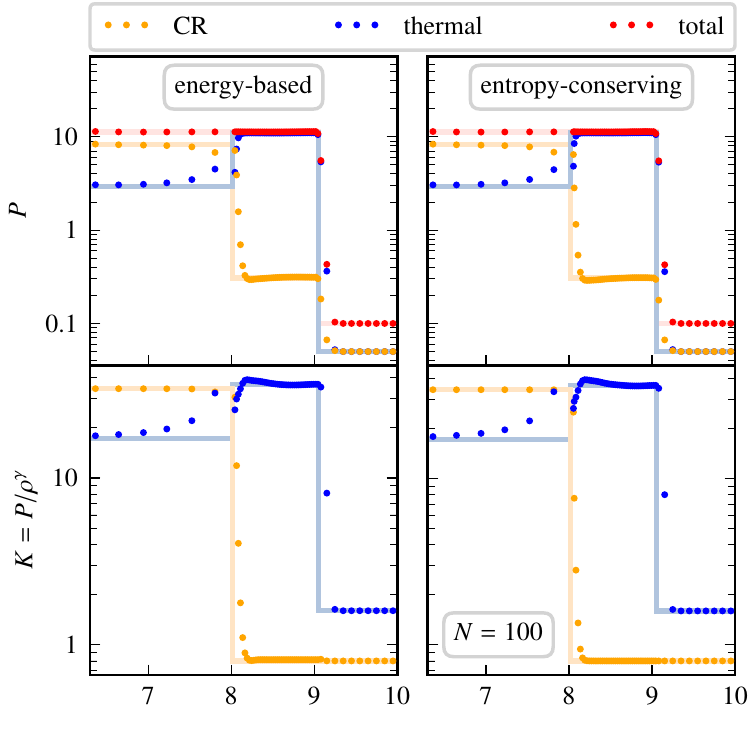}
    \includegraphics[width=0.49\linewidth]{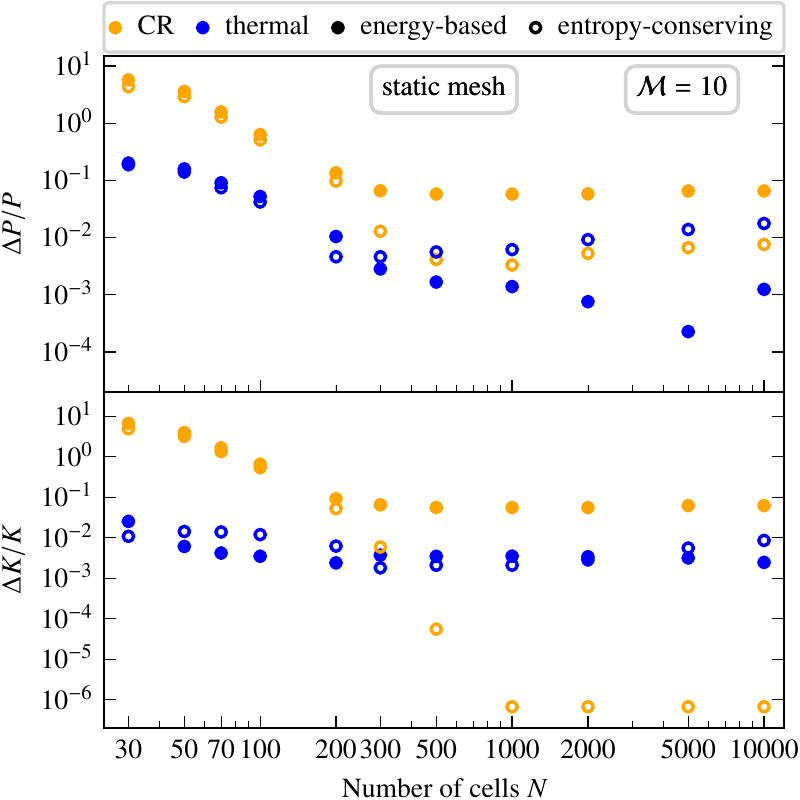}
    \includegraphics[width=0.49\linewidth]{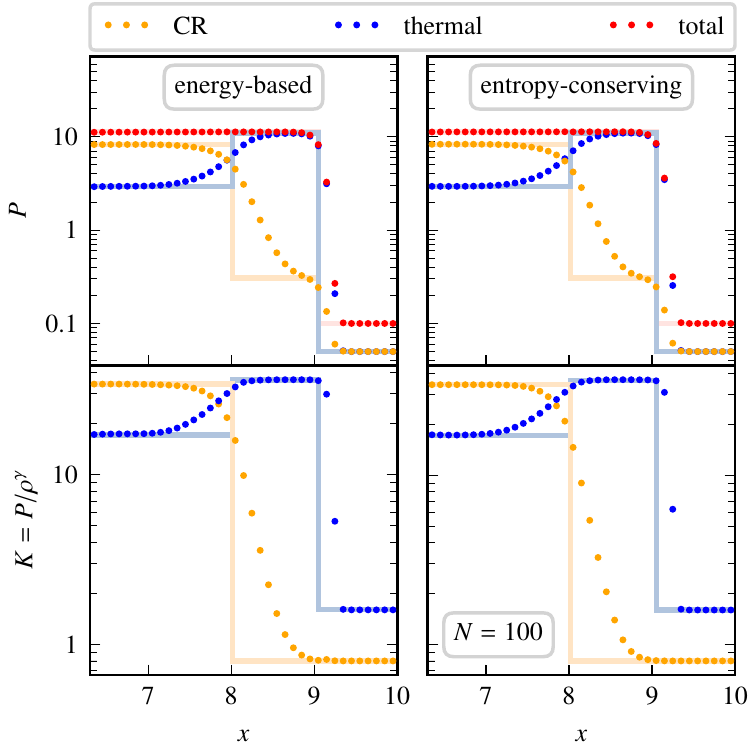}
    \caption{Shock-tube test with $\mathcal{M}=10$ and various resolutions ranging from $N=30$ to $10^4$ mesh cells \textit{without} accounting for CR acceleration at the shock. The top row displays the results of the moving-mesh approach, the bottom row shows the results using a static-mesh setup. In the left-hand panels, we plot the median of the absolute deviations in the post-shock region from the analytic solution of $P_{\rmn{th}}$, $P_{\rmn{cr}}$, $K_{\rmn{th}}$ and $K_{\rmn{cr}}$ at $t=0.37$. Filled circles indicate the results using the energy-based method, open circles indicate results of the entropy-conserving scheme. In each panel, we plot the relative error in $P$ in the top row, the relative error in $K$ in the bottom row. For a resolution of $N=100$ mesh cells, the corresponding post-shock region is depicted in the panels on the right-hand side, wherein the left column shows the results for the energy-based method and the right column for the entropy-conserving scheme. The static-mesh method yields significantly worse results due to its inherently higher numerical diffusivity.}
    \label{fig:errEntropyCRNumPart}
\end{figure*}

We analyse how this spurious entropy generation at the shock depends on the mesh resolution. To this end, we perform a sequence of test runs varying the number of mesh cells in the range of $N=30$ to $10^4$ while keeping the Mach number constant at $\mathcal{M}=10$. We run each simulation with both a moving mesh and the a fixed mesh to compare the two approaches. In order to quantify the deviation from the analytic solution, we evaluate the post-shock regime and determine the median of the absolute difference between the numerical and analytic solution within that region. We chose to calculate the median difference because the large entropy jump between the contact discontinuity and post-shock region would lead to misleading results when calculating the \textit{mean} deviation in low-resolution simulations.

In Fig.~\ref{fig:errEntropyCRNumPart}, we show the median differences of the thermal and CR pressures and entropy densities for varying resolutions from $N=30$ to $10^4$ on the left-hand side and display the pressure and entropy density profiles near the shock for $N=100$ on the right-hand side. Results obtained with the moving-mesh method are grouped together in the top row while the result obtained with the static-mesh method can be found in the bottom row.

In the static-mesh setup, the error in $K_{\rmn{cr}}$ diverges towards lower mesh resolutions for both the entropy- and energy-conserving numerical schemes which can be attributed to the higher numerical diffusivity of this approach. Only for a resolution of $N=200$ cells the deviations start to fall below 10 per cent and stabilizes towards higher resolutions or nearly vanishes for the entropy-conserving scheme. The behaviour of $P_{\rmn{cr}}$ is similar: while the error diverges in the poorly resolved runs for both methods, it stabilizes at around 7 per cent for the energy-based method and in the 1 per cent regime for the entropy-conserving scheme. The deviation of the thermal pressure $P_{\rmn{th}}$ is moderate for a small number of mesh cells and converges for higher resolutions to around $2$ per cent for the entropy-conserving scheme and to negligible values for the energy-based method. 

The moving-mesh approach consistently gives significantly better results. Even for very low resolutions, the deviation of $K_{\rmn{cr}}$ and $P_{\rmn{cr}}$ is clearly below 10 per cent for both energy- and entropy-conserving methods. In the high-resolution runs, these errors converge to around 2 per cent using the energy-based method, and nearly vanish when we apply the entropy-conserving scheme. The error in $K_\rmn{th}$ behaves nearly identical for both numerical schemes with values around 8 per cent for very low resolutions and negligible deviations for the high resolution runs. We find similar trends for $P_\rmn{th}$ but notice deviations in the 2 per cent regime for the lowest resolutions and negligible errors for an increasing number of mesh cells. In Appendix \ref{app:3Dshocktubes}, we demonstrate that the moving-mesh approach also yields appropriate results for a corresponding three-dimensional (3D) setup of the shock tubes.

We continue by investigating the dependency of spurious entropy generation at shocks on the Mach number $\mathcal{M}$. Again, we perform a suite of shock-tube simulations but fix the resolution at $N=100$ mesh cells and vary the Mach number in the range of $\mathcal{M}=1.5$ to $100$ this time. Shocks with lower Mach numbers require more time to fully develop. Hence, in each simulation, we evaluate the post-shock region once the shock has crossed $x=9$, which corresponds to the theoretical shock position at $t=0.37$ for $\mathcal{M}=10$ employed in the previous setup. Since we have already demonstrated that the moving-mesh setup gives much better results, we will stick to this approach in the following. 

\begin{figure}
    \includegraphics[width=\linewidth]{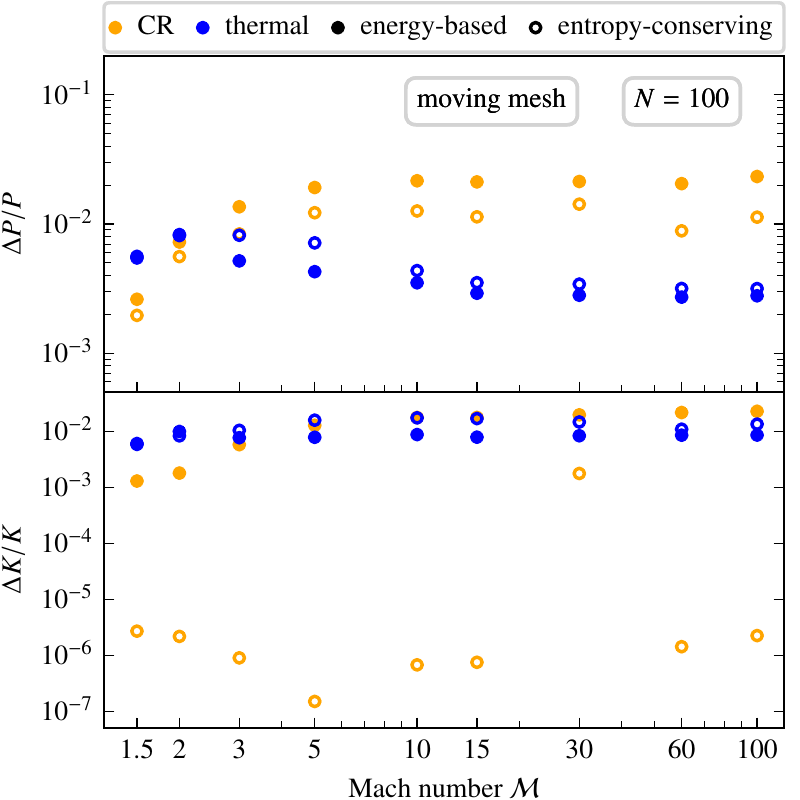}
    \caption{Shock-tube test with $N=100$ mesh cells and various Mach numbers ranging from $\mathcal{M}=1.5$ to $100$. We use our moving-mesh setup \textit{without} accounting for CR acceleration. We plot the median of the absolute deviations of $P_{\rmn{th}}$, $P_{\rmn{cr}}$, $K_{\rmn{th}}$ and $K_{\rmn{cr}}$ from the analytic solution in the post-shock regime. In each case, we evaluate the post-shock region once the shock has (theoretically) crossed $x=9$. Filled circles indicate the results of the energy-based method, open circles those of the entropy-conserving scheme. Shown are the relative error in $P$ in the top row and the relative error in $K$ in the bottom row.}
    \label{fig:errEntropyCRMach}
\end{figure}

Figure~\ref{fig:errEntropyCRMach} shows the results of the different runs. Again, the entropy-conserving scheme performs very good in adiabatically compressing the CRs at the shock with almost vanishing deviation in $K_\rmn{cr}$, independent of Mach number. The relative errors in $P_{\rmn{th}}$ and $P_{\rmn{cr}}$ slightly vary in the regime of $1$ per cent and remain small for higher Mach numbers. The energy-based method shows very similar results, except for the deviation of $K_{\rmn{cr}}$, which slightly increases up to a Mach number of $10$ and stabilizes at very small values of about $2$ per cent for larger $\mathcal{M}$. Overall, both methods give very good results and do not show a severe dependence on Mach number. \citet{Semenov2021} also performed the same test employing both the energy and entropy formalism for CR transport with the ART code. Using their implementation for the energy-based formulation of CR transport, they find a strong dependence of the CR entropy error on the Mach number with errors reaching $\lesssim20$ per cent for $\mathcal{M} \geq 9$.

\subsubsection{CR acceleration at the shock}
\label{subsub:CRAcc}
\begin{figure}
    \centering
    \includegraphics[width=\linewidth]{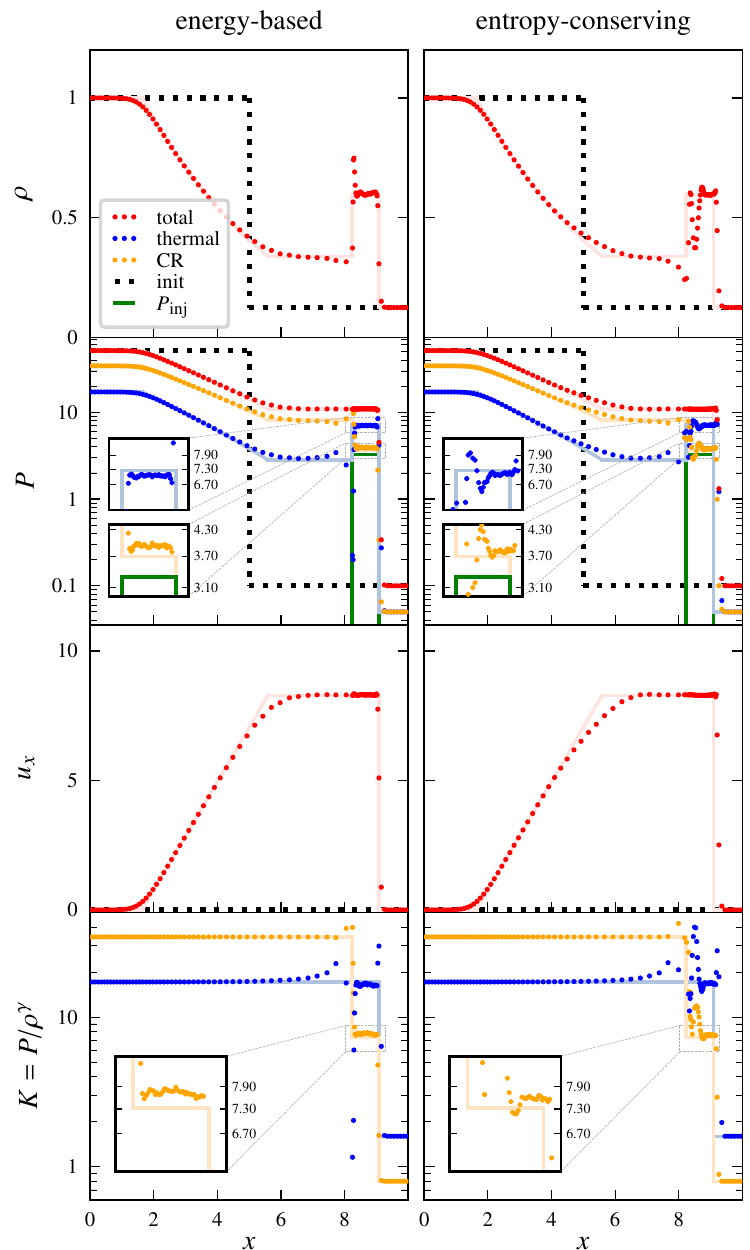}
    \caption{Same setup as in Fig.~\ref{fig:ShockTubeM10p100}, but now taking into account CR acceleration at the shock with $N=100$, $\mathcal{M}=9.56$ and the snapshot taken at $t=0.39$.}
    \label{fig:ShockTubeM10p100Acc}
\end{figure}

Figure~\ref{fig:ShockTubeM10p100Acc} shows the results of the 1D shock-tube test with $\mathcal{M}=9.56$ including CR acceleration at the shock. The left-hand panel displays the results that we obtained with the energy-based method and the right-hand panel shows the results using the entropy-conserving scheme. Again, we perform each run with identical initial conditions (cf. Table~\ref{table:inits}) and a spatial resolution of $N=100$ mesh cells.

\begin{figure}
    \centering
    \includegraphics[width=\linewidth]{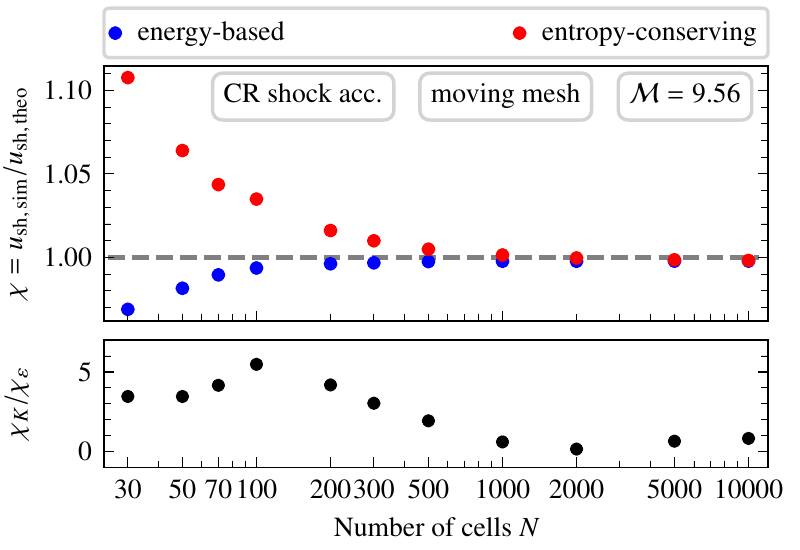}
    \caption{Ratio of simulated-to-theoretical shock velocity $\chi = u_\rmn{sh,\,sim} / u_\rmn{sh,\,theo}$ (top panel) and the $\chi$-ratio of both numerical schemes (bottom panel) as a function of resolution $N$, respectively. We use the moving-mesh setup of \textsc{Arepo} with a fixed Mach number of $\mathcal{M}=9.56$ and account for CR acceleration at the shock. Results obtained with the energy-based method (index $\varepsilon$) are coloured blue, those obtained with the entropy-conserving scheme (index $K$) red.}
    \label{fig:errorShockZone}
\end{figure}

The results obtained with energy-based method agree with the exact solution up to minor deviations. The most pronounced differences are the relatively high blips in density, pressure and entropy in the first two cells past the contact discontinuity. This comes about because in the first few time steps after the start of the simulation, when the shock has not yet fully developed and the post-shock regime is about to form, our algorithm injects too much CR energy because the estimated pressure jump is initially too large. While this causes an increased compressibility in comparison to the exact solution, the algorithm recovers as soon as the shock and post-shock regime have formed and then performs correctly. This behaviour was already mentioned in \citet{Pfrommer2017}. Zooming into the post-shock regime, we find that $K_\rmn{cr}$ and $P_\rmn{cr}$ are subject to a $\sim6$ per cent error, while $P_\rmn{th}$ deviates by 3 per cent. 

The entropy scheme, however, performs worse in this setup. Again, we notice the blips in density and entropy, but in the opposite direction. Unlike the energy-based method, these blips do not settle down when the post-shock zone has developed, but form oscillations with fairly large amplitudes that pervade half of the post-shock region. This is because CR entropy is injected at the shock and therefore CRs are not adiabatically compressed, making entropy conservation no longer valid and the algorithm has problems to adjust to the sudden change of the initially conserved quantity. Most importantly, the shock propagates to fast in comparison to the analytical solution in the entropy-based scheme. This is a consequence of mass conservation: because the density is too low in the left-hand part of the post-shock zone, the total post-shock zone needs to be broader and the shock advances faster. 

To quantify this behavior, we evaluate the ratio of the simulated-to-theoretical shock velocity $\chi = u_\rmn{sh,\,sim}/u_\rmn{sh,\,theo}$ by averaging 10 snapshots in the period from $t=0.31$ to $t=0.4$. In Fig.~\ref{fig:errorShockZone}, we plot the result as a function of resolution. Here, we use the moving-mesh setup, a fixed Mach number of $\mathcal{M}=9.56$, and we vary the resolution in the range of $N=30$ to $10^4$. The energy-based method simulates the shock position very accurately even for the lowest-resolution run, amounting to a deviation from the theoretical value of $\lesssim3$ per cent. The error quickly reaches negligible values for higher resolutions. Using the entropy-conserving scheme, the simulated shock position is significantly less accurate in comparison to the energy method, particularly for low resolutions, where the deviation is $\gtrsim 10$ per cent, more than four times worse in comparison to the energy-based method. Only for a resolution of $N\gtrsim500$ the entropy scheme approaches the accuracy of the energy-based method and the oscillations described earlier also vanish.

\begin{figure}
    \includegraphics[width=\linewidth]{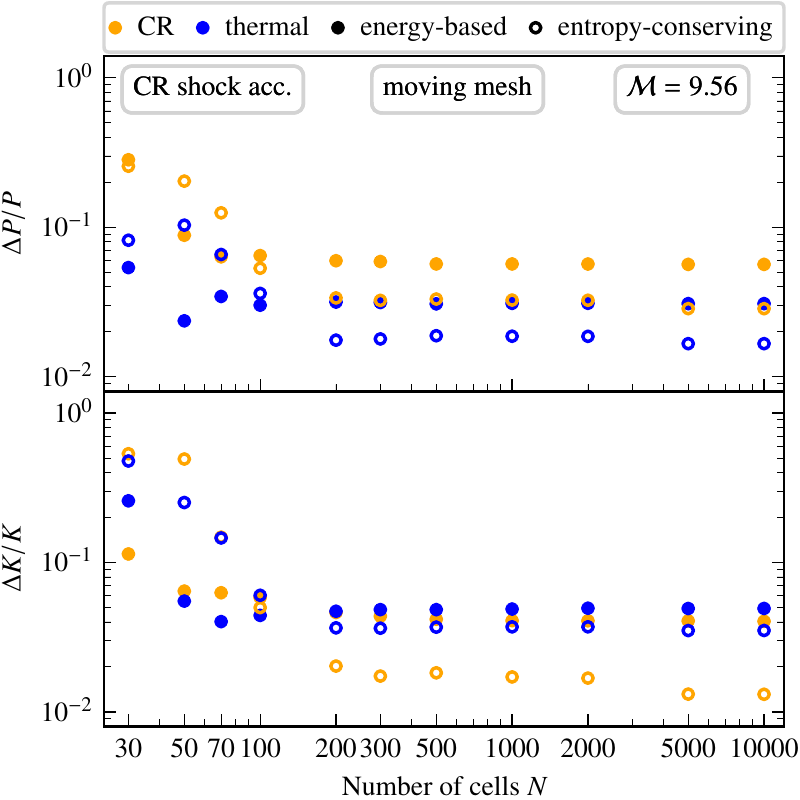}
    \caption{Shock-tube test with Mach number $\mathcal{M}=9.56$ and various resolutions ranging from $N=30$ to $10^4$. We use our moving-mesh setup and account for CR acceleration at shocks. We plot the median of the absolute deviations of $P_{\rmn{th}}$, $P_{\rmn{cr}}$, $K_{\rmn{th}}$ and $K_{\rmn{cr}}$ from the analytic solution in the post-shock regime at $t=0.39$. Filled circles indicate the results of the energy-based method, open circles those of the entropy-conserving scheme. Shown are the relative error in $P$ in the top row and the relative error in $K$ in the bottom row.}
    \label{fig:errEntropyCRinjNumPart}
\end{figure}

We investigate the dependence of the error on the number of mesh cells $N$ for our current setup that includes CR acceleration. Therefore, we fix the Mach number at $\mathcal{M}=9.56$ and vary the resolution in the range of $N=30$ to $10^4$. Figure~\ref{fig:errEntropyCRinjNumPart} shows the results of our test runs. As expected, the inclusion of CR acceleration worsens the numerical solution so that truncation errors at high resolution amount to about 6 per cent for $P_\rmn{cr}$ and 4 per cent for $K_\rmn{cr}$ (energy-based method) and approximately half of that for the entropy-conserving scheme. At low resolution, the errors increase to values exceeding 10 per cent, with the errors in the entropy-conserving scheme to rise above those in the energy-based method. Note that we identify the error with the median of the absolute deviation between simulation and theory so that the error is not sensitive to (even significant) post-shock oscillations as long as they do not accumulate to more than half of the mesh cells within the post-shock region. Because the oscillations are confined to only a few cells, the median error is hence only slightly affected by this feature while we identified it to have a significant impact on the shock propagation at resolutions $N\lesssim500$ (see Fig.~\ref{fig:errorShockZone}).

\section{Isolated models of galaxy formation}
\label{sec:isolatedmodelsofgalaxyformation}
In this section, we continue our comparison of the energy-based method and the entropy-conserving scheme in a more realistic astrophysical application. We simulate the formation of three different isolated galaxies inside halo masses of $10^{10}$, $10^{11}$ and $10^{12} \, \rmn{M}_\odot$.

\begin{figure*}
    \includegraphics[width=0.495\linewidth]{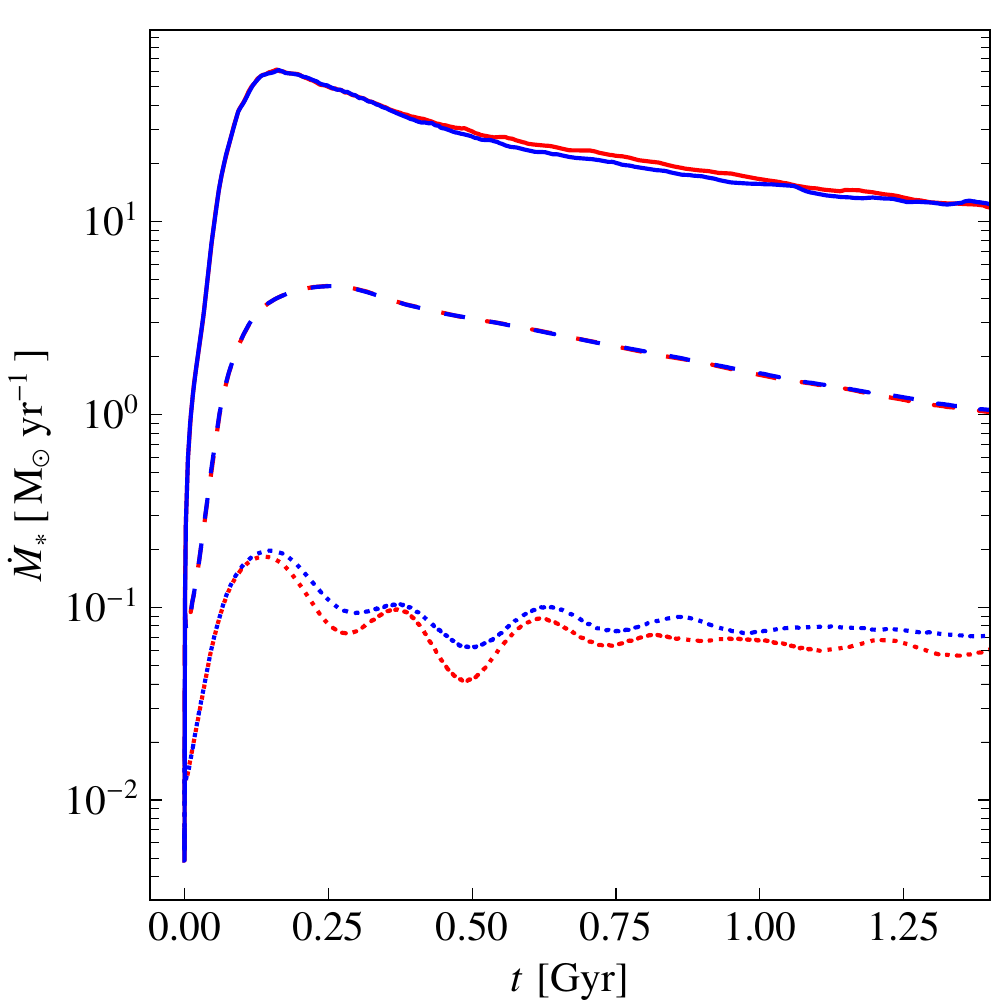}
    \includegraphics[width=0.495\linewidth]{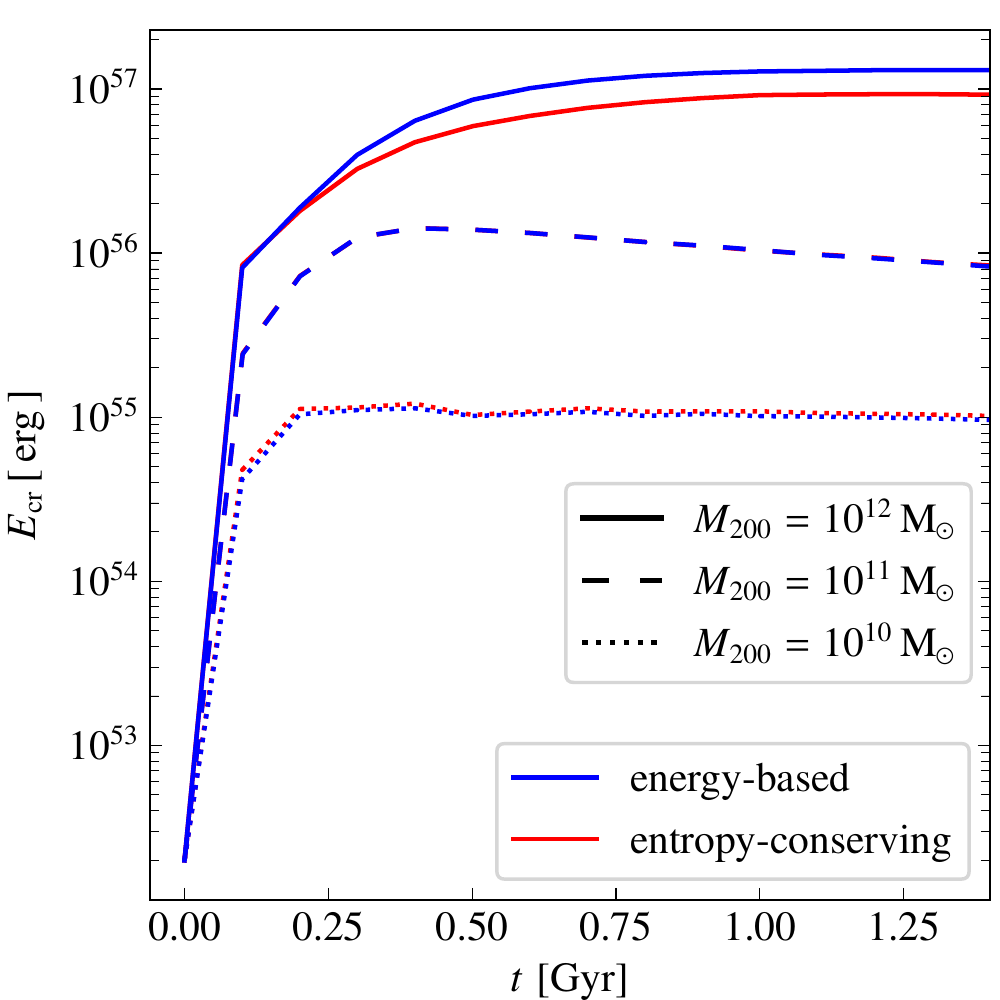}
    \caption{Star-formation rate (SFR, left-hand panel) and instantaneous CR energy (right-hand panel) as a function of time for our various haloes. Profiles for these quantities are depicted using dotted, dashed and solid lines for the haloes of mass $10^{10}$, $10^{11}$ and $10^{12} \, \rmn{M_\odot}$, respectively. Results using the energy-based method are coloured blue, those of the entropy-conserving scheme red.}
    \label{fig:SFR_ECR}
\end{figure*}

We model the interstellar medium (ISM) by an effective pressurised equation of state and follow radiative cooling and star formation using the approach by \citet{Springel2003}. In addition to the composite thermal and CR fluid, we evolve the magnetic field using the  \citet{Powell1999} scheme for divergence control as implemented in \textsc{Arepo} \citep{Pakmor2013}. The magnetic field is initialised with a low-amplitude uniform seed magnetic field with a strength of $B=10^{-10} \, \rmn{G}$. The general setup is identical to the one used in \citet{Pfrommer2017}. We adopt Navarro-Frenk-White (NFW) profiles for the dark matter component \citep{Navarro1997} which are characterized by the concentration parameter $c_{200}=r_{200}/r_\rmn{s}$ where $r_{200}$ denotes the radius that encloses 200 times the critical density of the universe and $r_\rmn{s}$ is the characteristic radius of the NFW profile. We chose the values for $c_{200}$ following the results presented by \citet{Maccio2008}. We adopt a hydrostatic gas distribution that is initially in equilibrium within the halo. We assume that the halo carries a small amount of angular momentum, parametrized by a spin parameter $\lambda=J|E|^{1/2}\rmn{G}^{-1}M_{200}^{-5/2}$, where $J$ is the angular momentum, $|E|$ is the total halo energy, $\rmn{G}$ is the gravitational constant and $M_{200}$ denotes the mass within $r_{200}$. For each run we chose $\lambda=0.05$ and a baryon mass fraction of $\Omega_\rmn{b} / \Omega_\rmn{m} = 0.155$. 

In the initial conditions of our high-resolution simulations, we have $N=10^7$ gas cells inside the virial radius. Each gas cell has a mass of $155 \rmn{M_\odot} \times  M_{200}/(10^{10}\rmn{M}_\odot)$ which also corresponds to the target mass of the cells throughout the simulation. We enforce that the mass of all cells is within a factor of 2 of the target mass by explicitly refining and de-refining the mesh cells that violate these criteria. We additionally require that adjacent cells adhere a maximum volume difference (MVD) of 10 and refine the larger cell if this condition is violated. Furthermore, we adopt a threshold for the star-forming density of $\rho_\rmn{sf}=5.98\times 10^{-3} \, \rmn{M_\odot \, pc^{-3}}$. We account for CR injection at SNe with a CR energy injection efficiency of $\zeta_\rmn{SN}=0.1$ which indicates the fraction of SN energy that is converted into CRs. The CR injection at SNe is performed with a sub-resolution model and not with our explicit shock finding method and associated CR acceleration.\footnote{For a detailed description of the sub-resolution model, we refer to Section 3.2 in \citet{Pfrommer2017}.} We assume advective CR transport and account for adiabatic changes of the CR energy as well as Coulomb and hadronic CR cooling, while neglecting active CR transport in form of anisotropic diffusion and streaming. A summary of the simulation parameters is listed in Table~\ref{tab:galaxyinits}.

\begin{table}
    \caption{Parameters of the isolated galaxy simulations. Columns from left to right label (1) virial mass $M_{200}$, (2) concentration parameter of the NFW profile, (3) initial gas fraction, (4) dimensionless spin parameter, (5) CR acceleration efficiency at SNe, (6) initial number of resolution elements $N$ within the virial radius, and (7) maximum volume difference (MVD) of adjacent Voronoi cells.}
    \centering
    \begin{tabular}{ccccccc}
        \hline\hline
        $M_{200}$ & $c_{200}$ & $\Omega_\rmn{b} / \Omega_\rmn{m}$ & $\lambda$ & $\zeta_\rmn{SN}$ & $N$ & $\rmn{MVD}$\\
        \hline
        $10^{10} \, \rmn{M}_\odot$ & 11  &  0.155 & 0.05 & 0.1 & $10^{7}$ & 5, 10\\
        $10^{11} \, \rmn{M}_\odot$ & 8.5 &  0.155 & 0.05 & 0.1 & $10^{7}$ & 10\\
        $10^{12} \, \rmn{M}_\odot$ & 7   &  0.155 & 0.05 & 0.1 & $10^{7}$ & 5, 10\\
        $10^{12} \, \rmn{M}_\odot$ & 7   &  0.155 & 0.05 & 0.1 & $10^{6}$ & 10\\
        $10^{12} \, \rmn{M}_\odot$ & 7   &  0.155 & 0.05 & 0.1 & $10^{5}$ & 10\\
    \end{tabular}
    \label{tab:galaxyinits}
\end{table}

In Fig.~\ref{fig:SFR_ECR}, we plot the star-formation rate (SFR, left-hand panel) and the instantaneous CR energy (right-hand panel) as a function of time for our three different haloes (shown with different line styles). Results using the energy-based method are coloured blue, those of the entropy-conserving scheme are shown with red. The $10^{10} \, \rmn{M}_\odot$ halo shows a slightly but systematically lower SFR using the entropy-conserving scheme, which can be explained by the minor increase in the corresponding CR energy. In comparison to the energy-based method, the higher pressure induced by CRs causes the thermal gas to cool more slowly which in turn leads to a decrease in the SFR. This effect declines with increasing halo mass as already shown by \citet{Pfrommer2017}. Hence, the same but opposite behaviour can be analogously explained for our $10^{12} \, \rmn{M}_\odot$ halo where the total CR energy is reduced by about 30 per cent using the entropy-conserving scheme. This leads to a small increase in SFR in the period between 0.5 and 1.2 Gyr. The halo with $10^{11} \, \rmn{M}_\odot$ shows no differences at all, neither in SFR nor in CR energy. 
We explain the behaviour for the various haloes as follows. Because the entropy-conserving scheme does not explicitly conserve CR \textit{energy}, this scheme introduces intrinsic differences in the CR energy when we compare it to the energy-based method. Thus, the temporal evolution of the CR energy for both schemes inevitable deviates. This leads to discrepancies in the SFR which in turn changes the amount of CRs injected. Thus, a cycle of altered CR energy is created in which the injection and non-conservation of CR energy influence each other through their effects on the SFR.\footnote{Another point that should not be ignored is the fact that the gain and loss terms in equation (\ref{eq:crenergy}) and (\ref{eq:crentropy}) describe variations in energy, not entropy. While the algebraic conversion of this is straightforward, the underlying physics may not be so easily transferable and should therefore be used with caution.} 

\begin{figure*}
    \includegraphics[width=\linewidth]{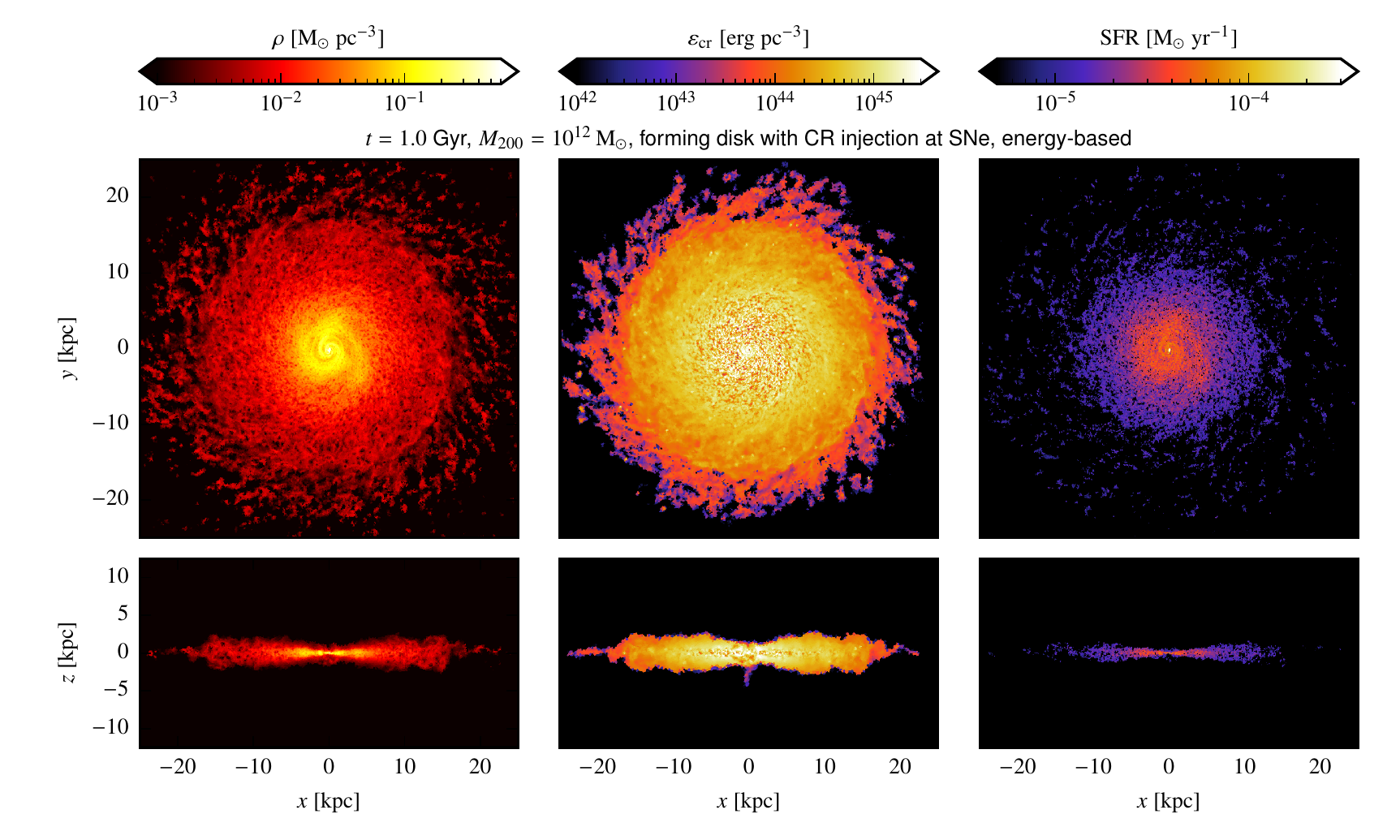}
    \includegraphics[width=\linewidth]{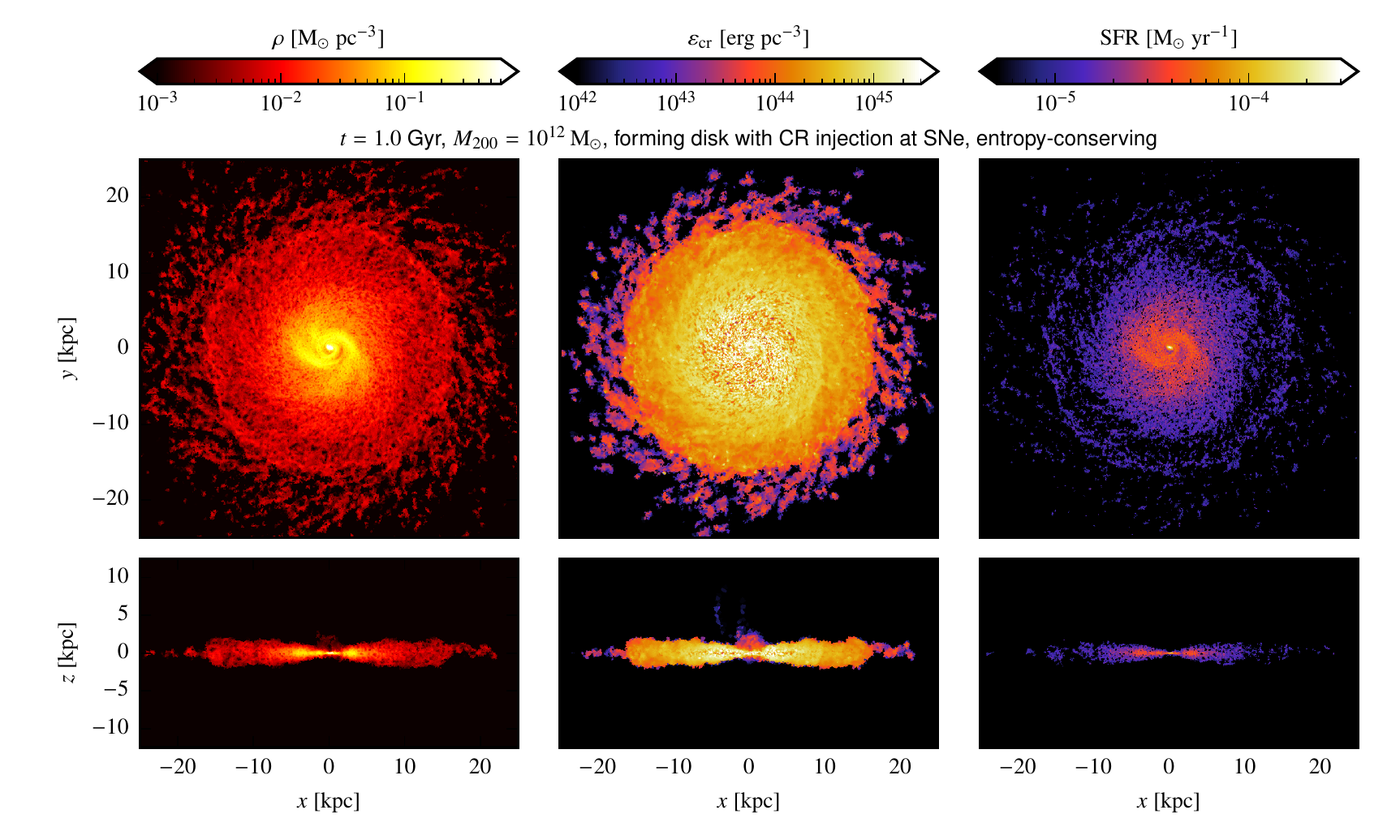}    
    \caption{Slices showing the gas density $\rho$, CR energy density $\varepsilon_\rmn{cr}$ and SFR (from left to right) for the galaxy situated in the $10^{12} \, \rmn{M_{\odot}}$ halo after 1 Gyr of evolution. The top six panels show results obtained with the energy-based method and the bottom six panels show the results of the simulation that employs the entropy-conserving scheme. For each quantity, we show slices through the mid-plane of the disc (face-on views) and vertical slices through the centre (edge-on views).
}
    \label{fig:galaxyDiscs}
\end{figure*}

In Fig.~\ref{fig:galaxyDiscs}, we show a gallery of slices that display the gas density $\rho$, CR energy density $\varepsilon_{\rmn{cr}}$, and SFR for the $10^{12} \, \rmn{M}_\odot$ halo after 1 Gyr of evolution. The top six panels depict the results using the energy-based method, the bottom six panels the results from the entropy-conserving scheme. Both numerical methods produce very similar results. At this stage of evolution, gas has rapidly accumulated in the centre of the galaxy, which leads to an increased gas density and SFR there. Most CRs are injected in this area as confirmed by the centrally enhanced CR energy density (panels in the middle row). While the distribution of the gas density in both haloes looks almost identical, the edge-on views of $\varepsilon_{\rmn{cr}}$ (bottom panels in the middle row) show a slightly more extended distribution of CR energy when the energy-based method is used. This is due to the increased CR pressure (or CR energy, cf.\ right panel in Fig.~\ref{fig:SFR_ECR}) providing additional pressure support. Furthermore, we notice a minor increase in SFR within a ring at about 14 to 16~kpc from the centre when using the entropy-conserving scheme. This is in agreement with a moderately reduced $\varepsilon_{\rmn{cr}}$ in this region in comparison to the energy-based method, as discussed in the previous paragraph. However, we note that the differences are minuscule and that the overall morphological appearances of both galaxies are nearly identical, especially considering the larger astrophysical uncertainties of the adopted model parameters. We demonstrate in Appendix~\ref{app:refinementcriterion} that the observed and already-small statistical differences can be further reduced if we adopt a more aggressive mesh-regularization strategy in the high-resolution simulation runs with initially $N=10^7$ mesh cells within the virial radius and we show in Appendix~\ref{app:convergence} approximate numerical convergence for the simulation of the $10^{12} \, \rmn{M}_\odot$ halo.

Note that recent galaxy simulations by \citet{Semenov2021} find larger differences between the entropy-conserving and energy-based methods. The main differences in comparison to our approach are their employed hydrodynamical method (a spatially fixed, adaptively refined mesh) and their explicitly modelled multi-phase ISM while we adopt an effective equation of state that results in a smoother ISM. \citet{Semenov2021} follow the radiative cooling down to temperatures of 40~K so that energy deposition into the cooling phase by supernovae result in more compressible, radiative shocks. Studying CR acceleration at radiative shocks is beyond the scope of this work and will be postponed to future work.

\section{Conclusions}
\label{sec:conclusion}

Here, we study various approaches to integrate CRs into MHD simulations, namely the energy-based method and the entropy-conserving scheme, in the context of the moving-mesh code \textsc{Arepo}. To this end, we perform a sequence of 1D shock-tube tests, with and without accounting for CR acceleration at shocks as well as using a static-mesh and a moving-mesh setup. This allows us to analyse the idealized behavior of CRs under the influence of adiabatic and non-adiabatic changes using different numerical schemes, in addition to comparing the performance of the two mesh approaches. Moreover, we use both numerical methods to simulate the influence of CRs on the formation of several isolated galaxies in haloes of mass $10^{10}$, $10^{11}$ and $10^{12} \, \rmn{M_{\odot}}$ including advective CR transport and feedback in terms of CR injection by SNe. We find that:

\begin{itemize}
    \item The moving-mesh approach performs significantly better than the static-mesh setup, which is due to the comparably high numerical diffusivity of the latter. This is true regardless of the method used to integrate the CRs (see Fig.~\ref{fig:errEntropyCRNumPart}).
    \item At \textit{very high resolution}, the entropy-conserving scheme has a lower error in CR energy by a factor of 10 when omitting CR acceleration (cf. top row in Fig.~\ref{fig:errEntropyCRNumPart}) and by a factor of 2 when accounting for CR acceleration at shocks (see Fig.~\ref{fig:errEntropyCRinjNumPart}). However, the overall error remains small (less than 2 per cent and 6 per cent, respectively) for the energy-based method and hence far below astrophysical uncertainties.
    \item At \textit{low resolution}, which is more typical for astrophysical large-scale simulations, both numerical schemes perform almost identical in terms of CR and thermal energy in a setup without CR acceleration (see Fig.~\ref{fig:errEntropyCRNumPart}). When considering CR acceleration at the shock, the energy-based method proves to be numerically much more stable (see Fig.~\ref{fig:ShockTubeM10p100Acc}) and thus shows significantly lower deviations from the analytic solutions, particularly in CR entropy (see Fig.~\ref{fig:errEntropyCRinjNumPart}).
    \item The shock velocity is determined significantly more accurately using the energy-based method when CR acceleration at the shock is considered, particularly at low and intermediate resolutions where deviations are reduced by a factor of 5 to 6 in comparison to the entropy-conserving scheme (see Fig.~\ref{fig:errorShockZone}).
    \item The simulations of isolated galaxies yield almost identical results using either numerical method (see Fig.~\ref{fig:galaxyDiscs}). The small variations in SFR and instantaneous CR energy (see Fig.~\ref{fig:SFR_ECR}) can be explained by the intrinsic behavior of the entropy-conserving scheme where energy is not explicitly conserved. 
\end{itemize}

In this work, we have demonstrated that the integration of CRs into MHD simulations using a moving-mesh approach can be properly achieved with either the energy-based method or the entropy-conserving scheme, as long as active CR acceleration at shocks is omitted. When the latter is considered, the energy-based method is the preferred choice, in particular for poorly resolved simulations.

\section*{Acknowledgements}

We thank Vadim Semenov and Andrey Kravtsov for constructive comments on the manuscript and acknowledge support by the European Research Council under ERC-CoG grant CRAGSMAN-646955 and ERC-AdG grant PICOGAL-101019746.

\section*{Data Availability}
The data underlying this article will be shared on reasonable request to the corresponding author.



\bibliographystyle{mnras}
\bibliography{references}


\appendix
\section{3D shock tubes}
\label{app:3Dshocktubes}

The 1D shock-tube test, described in Section~\ref{sec:testproblems}, is a useful tool for evaluating the general performance of a numerical method in an idealized environment. Here, we analyse the differences of the energy-based method and the entropy-conserving scheme in the more challenging 3D shock-tube setup. To address this, we set up a box of size $(L_x, L_y, L_z)=(10,1,1)$ and use an irregular glass-like distribution of the particles as initial conditions \citep[see][for details]{Schaal2015}. Like in the 1D case, we fix the Mach number at $\mathcal{M}=10$ and omit CR acceleration at the shock. We vary the number of mesh cells along the $x$ axis in $N_x=\{30, 50, 70, 100, 200\}$ and choose the number of mesh-generating points in the $y$ and $z$ direction to be $N_y=N_z=N_x/10$. 

\begin{figure}
    \centering
    \includegraphics[width=\columnwidth]{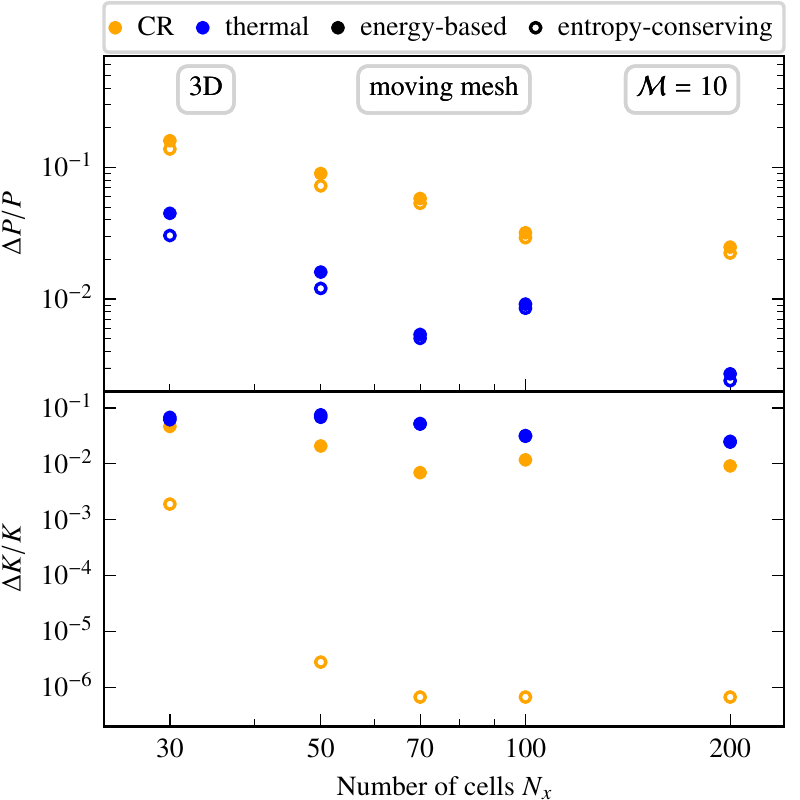}
    \caption{Same representation as in the upper left panel in Fig.~\ref{fig:errEntropyCRNumPart}, but with a 3D setup. $N_x$ denotes the number of mesh cells along the $x$-axis, and the $y$ and $z$ resolutions are chosen to be $N_y = N_z = N_x/10$, respectively. }
    \label{fig:3Dshocktubes}
\end{figure}

We apply the same statistical analysis as in Section~\ref{sec:testproblems} and plot the median absolute deviation of the simulation result from the analytic solution in  Fig.~\ref{fig:3Dshocktubes}. The trend of these errors is similar to the one obtained in 1D and shows that deviations get smaller for increased resolutions until they saturate at the 3-percent level. Interestingly, the pressure deviations do not differ significantly between the simulations employing the energy- or entropy-conserving scheme. However, the errors calculated for the three-dimensional simulations are larger if we directly compare them to those obtained from the corresponding 1D shock tube at the same resolution.

\section{Scrutinising numerical convergence in galaxy simulations}

\subsection{Adapting the refinement criterion}
\label{app:refinementcriterion}

\begin{figure*}
    \centering
    \includegraphics[width=0.49\linewidth]{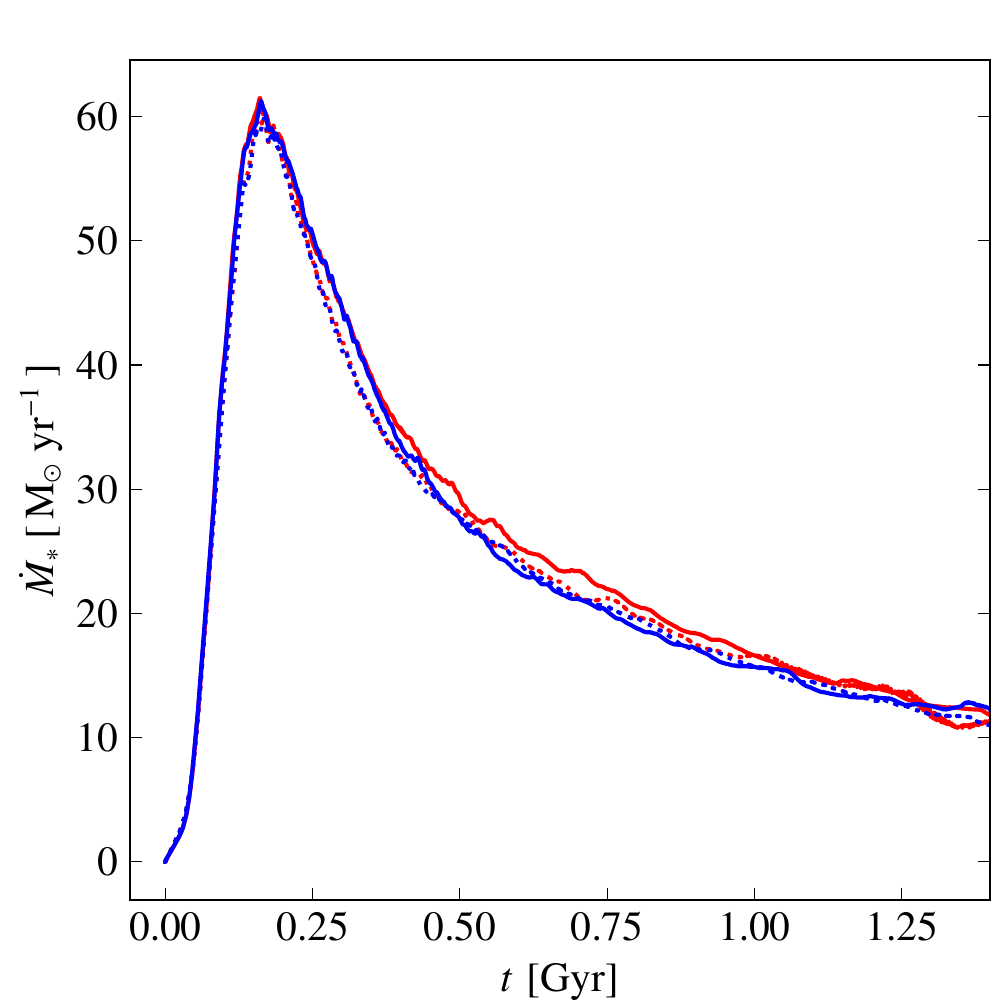}
    \includegraphics[width=0.49\linewidth]{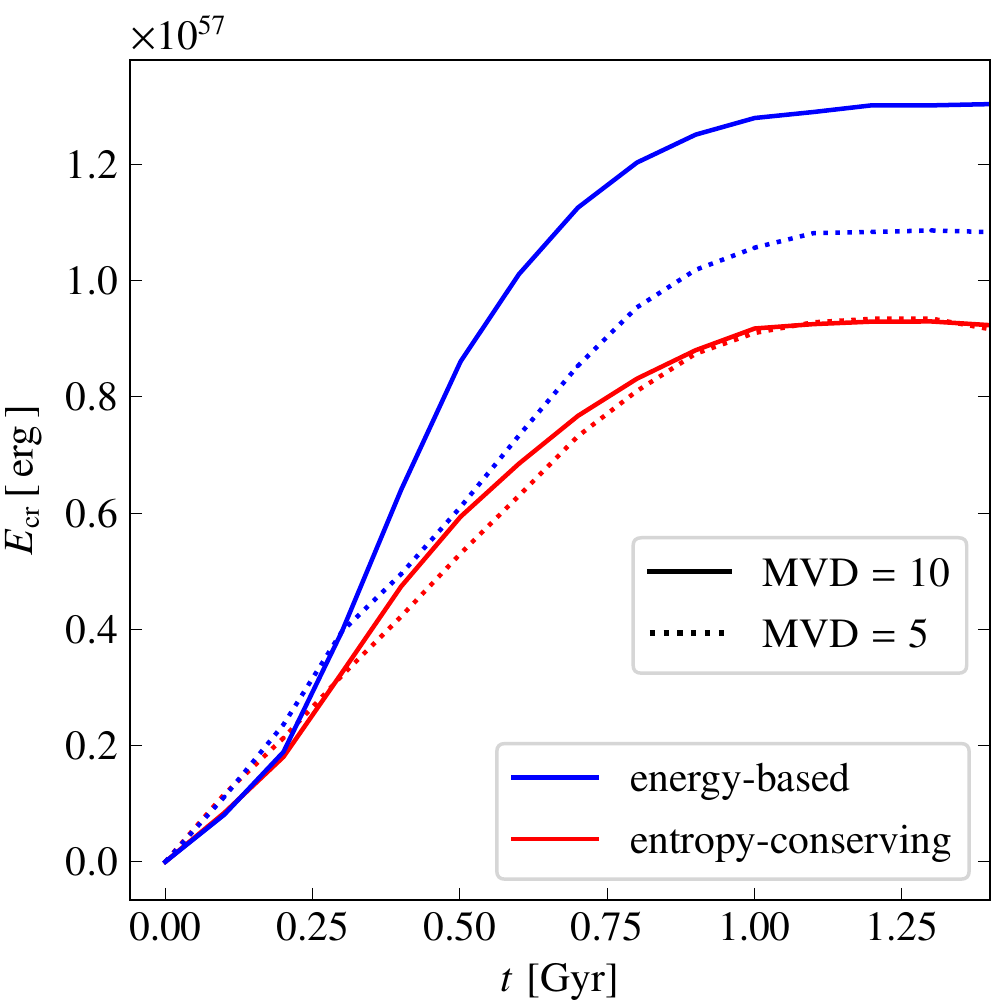}
    \caption{SFR (left panel) and instantaneous CR energy within the simulation (right panel) as a function of time for our halo with $10^{12} \, \rmn{M}_\odot$ plotted with a linear scaling. Results using the energy-based method are coloured blue, those of the entropy-conserving scheme are shown in red. The solid lines show the results when neighboring cells differ in volume by a maximum factor of $10$, and the dotted lines show the results for a MVD of $5$.}
    \label{fig:SFR_MV5}
\end{figure*}

As stated in Section~\ref{sec:isolatedmodelsofgalaxyformation}, we limit adjacent cells to differ in volume at most by a factor of $10$ in our simulations of isolated galaxies. Here, we analyse a setup where this maximum volume difference (MVD) is restricted to a factor of $5$ which has the effect of resolving regions of high density even more accurately. This is of particular interest in terms of star formation and CR injection. In Fig.~\ref{fig:SFR_MV5}, we plot the SFR (left-hand panel) and instantaneous CR energy (right-hand panel) of the $10^{12} \, \rmn{M}_\odot$ halo and compare simulations with the fiducial and the more restrictive value for the MVD. Results using the energy-based method are coloured blue, those of the entropy-conserving scheme are shown in red. Solid (dotted) lines indicate the previous results using a MVD of $10$ (5). Both methods yield a very similar SFR. The instantaneous CR energy echos this finding, with the more restrictive MVD simulations to differ at most by less than 20 percent. Analogously, we adapt the refinement criterion for the $10^{10} \, \rmn{M}_\odot$ halo, but notice no change from our fiducial case with an MVD of 10.

\subsection{Convergence behaviour of numerical schemes}
\label{app:convergence}
\begin{figure}
    \centering
    \includegraphics[width=\columnwidth]{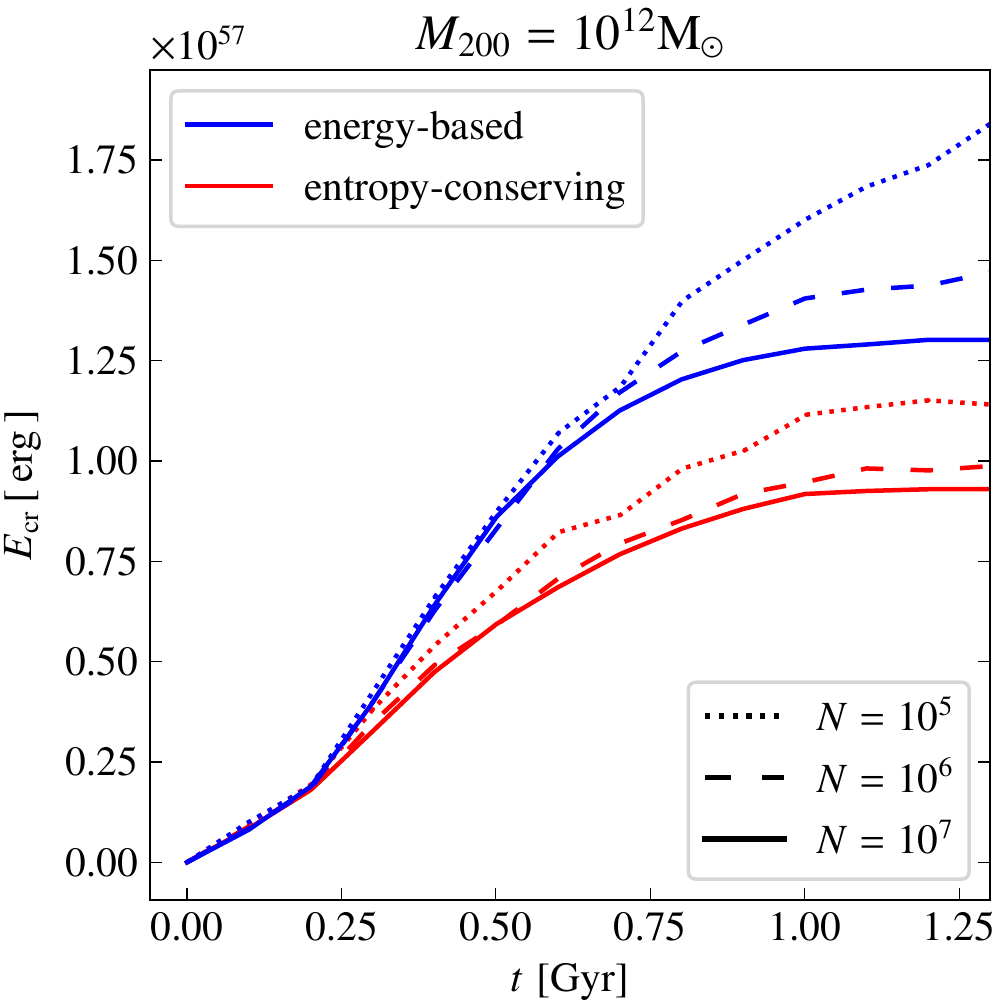}
    \caption{Total instantaneous CR energy as a function of time for different resolutions of the isolated galaxy simulations. Results using the energy-based method are coloured blue, outcomes of the entropy-conserving scheme red. Results of the runs with a resolution of $10^{5}$, $10^{6}$ and $10^{7}$ mesh cells are shown as dotted, dashed and solid lines, respectively.}
    \label{fig:convergence}
\end{figure}

In this Appendix, we show the convergence behavior of the energy-based and entropy-conserving methods as a function of resolution. To this end, we show the total CR energy of our $10^{12} \, \rmn{M}_\odot$ halo in Fig.~\ref{fig:convergence} and plot the results for initial resolutions of $10^{5}$, $10^{6}$, and $10^{7}$ grid cells as dotted, dashed, and solid lines, respectively. Results of using the energy-based method are coloured blue, results of the entropy-conserving scheme with red. We use an MVD of 10 in each case. Either method converges with an increasing number of mesh cells, albeit to different values, with the discrepancy between the two schemes decreasing with increasing resolution.


\bsp	
\label{lastpage}
\end{document}